\documentclass[%
 preprint,
 amsmath,amssymb,
 aps, physrev,
]{revtex4-2}

\usepackage{graphicx}
\usepackage{dcolumn}
\usepackage{bm}
\usepackage{physics}
\usepackage{xcolor}
\usepackage{bm}
\usepackage{xtab,afterpage,longtable}
\usepackage{lipsum}
\usepackage{graphicx}
\usepackage{dcolumn}
\usepackage{booktabs} 
\usepackage{multirow}
\usepackage{color}
\usepackage{eucal}
\usepackage{mathrsfs}
\usepackage{siunitx}
\usepackage{makecell}
\usepackage{hyperref}

\renewcommand{\d}{{\mathrm{d}}}

\renewcommand{\P}{\mathcal{P}}
\newcommand{\T}{\mathcal{T}}

\renewcommand{\AA}{\si{\angstrom}}

\begin{document}

\preprint{Materials Today Quantum}

\title{Linear and Nonlinear Edelstein Effects \\ in Chiral Topological Semimetals}%

\author{Haowei Xu}
\affiliation{%
 Department of Nuclear Science and Engineering, Massachusetts Institute of Technology, Cambridge, MA 02139, USA
}%

\author{Ju Li}%
 \email{liju@mit.edu}
 \affiliation{Department of Materials Science and Engineering, Massachusetts Institute of Technology, MA 02139, USA}
\affiliation{
   Department of Nuclear Science and Engineering, Massachusetts Institute of Technology, Cambridge, MA 02139, USA}

\date{\today}

\begin{abstract}
Recently, there has been growing interest in achieving on-demand control of magnetism through electrical and optical means. In this work, we provide first-principles predictions for the linear and nonlinear Edelstein effects (LEE and NLEE) in the chiral topological semimetal CoSi. The LEE and NLEE represent first- and second-order magnetic responses to external electric fields, enabling precise manipulation of magnetization via electrical and optical methods. We demonstrate that although both LEE and NLEE require time-reversal symmetry breaking, they can still be realized in non-magnetic materials, as time-reversal symmetry can be spontaneously broken by heat and dissipation, according to the second law of thermodynamics. Meanwhile, due to different inversion symmetry selection rules, the LEE and NLEE manifest opposite and identical signs in the two enantiomers of CoSi, respectively. We further quantify the magnitude of LEE and NLEE, showing that electrically or optically induced magnetization can reach 10 Bohr magneton per unit cell when the external electric field strength is comparable with the internal atomic electric field, which is on the order of $1~\rm V/\AA$. Our work offers a systematical approach for predicting the electrical and optical control of magnetism in real materials, paving the way for potential applications in areas such as spintronics and magnetic memories.
\end{abstract}

\maketitle


\section{Introduction}

Many applications, such as magnetic memory and spintronics~\cite{vzutic2004spintronics,bader2010spintronics}, require high-speed, precise control over magnetism. Traditionally, magnetism is controlled using external magnetic fields. However, generating magnetic fields on the Tesla scale, which is necessary for manipulating the magnetism of certain materials, presents significant challenges. Additionally, it is rather difficult to focus magnetic fields, hindering the development of functional miniaturized devices. In recent years, electrical and optical approaches have emerged as versatile alternatives for on-demand control over magnetism. The electrical control over magnetism with electric fields and/or associated spin \emph{unpolarized} electric currents can be realized with e.g., the magnetoelectric~\cite{fiebig2005revival,eerenstein2006multiferroic,huang2018electrical,iniguez2008first,malashevich2012full,scaramucci2012linear,ghiasi2019charge,el2023observation} and the linear Edelstein~\cite{edelstein1990spin} effects. Meanwhile, the optical approach has also attracted considerable attention as it can be non-contact, ultra-fast, and with high spatial resolution. Experimental demonstrations of optical control over magnetism have been reported in several prior studies~\cite{stanciu2007all,lambert2014all,mangin2014engineered,el2016two,van1965optically,wu2024reversible,zhang2022all,fang2024perspectives}, with mechanisms involving either thermal or non-thermal effects.

\begin{figure}
    \centering
    \includegraphics[width=0.75\linewidth]{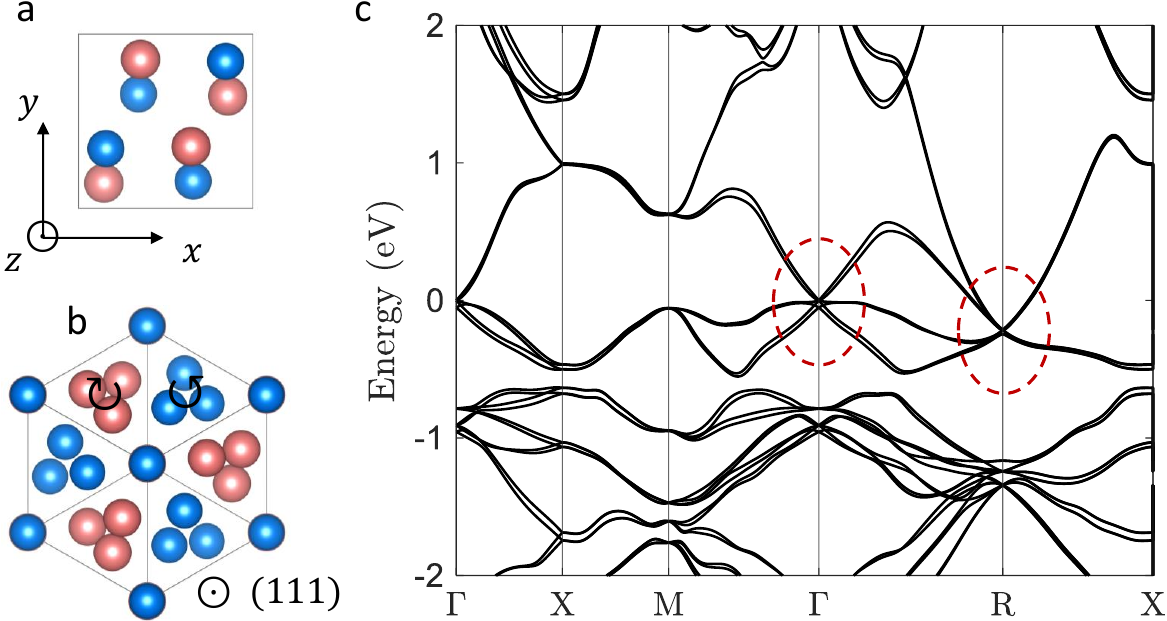}
    \caption{(a, b) Atomic structure of left-handed CoSi. The Co and Si atoms form a helical structure along the (111) direction, as denoted by curved arrows in (b). Pink: Co; Blue: Si. (c) Electronic band structure of CoSi along high-symmetry lines. The dashed circles mark the topological nodes at $\Gamma$ and $\rm R$ points in the Brillouin zone.}
    \label{fig:atomic_band_structure}
\end{figure}

While theoretical and simulation studies on various types of transport and optical properties are abundant, first-principles predictions regarding electrical and optical control of magnetism remain relatively scarce. In this work, we present a first-principles study on the linear~\cite{edelstein1990spin} and nonlinear~\cite{xu2021light} Edelstein effect (LEE and NLEE), which are the first- and second-order responses to external electric fields, and serve as generic mechanisms to generate static magnetization by electric and light fields, respectively. The LEE and NLEE effects are universal and apply to a wide range of materials.
Different from the permanent magnetic moments in magnetic materials, which are mostly stabilized by spin-spin interactions, the magnetization induced by LEE and NLEE are out-of-equilibrium properties that exist only when the electric/optical field is turned on.  We will use the chiral topological semimetal, CoSi~\cite{chang2018topological,tang2017multiple,xu2020optical,yan2024structural} as an example. CoSi features a chiral atomic structure (Figure~\ref{fig:atomic_band_structure}a), and its band electronic structure hosts Weyl nodes with non-zero topological charges at high-symmetry points of the Brillouin zone (Figure~\ref{fig:atomic_band_structure}b).

We will elucidate the symmetry requirements for LEE and NLEE. Both effects require time-reversal symmetry breaking, although they can still occur in non-magnetic materials, as the time-reversal symmetry can be spontaneously broken by heat and dissipation associated with charge current and/or light illumination. This resembles the chirality-induced spin selectivity, where it is argued that non-unitary processes such as dephasing and leakage break time-reversal symmetry~\cite{guo2014spin,matityahu2016spin}. Additionally, LEE necessitates inversion symmetry breaking and has opposite signs in the two enantiomers of CoSi (i.e. left- and right-handed structures). In contrast, NLEE is not restricted by inversion symmetry and would be identical in the two enantiomers. The strength of the LEE and NLEE scales linearly and quadratically with the electric field strength. In particular, we find that for both LEE and NLEE, the induced magnetization can reach $10~\rm \mu_{\rm B}$ (with $\mu_{\rm B}$ the Bohr magneton) when the electric field is of the order of $1~\rm V/\AA$. In the following sections, we will discuss the symmetry requirements, microscopic mechanisms, and numerical results for the LEE and NLEE in detail.

\section{Linear Edelstein effect}

We first examine the LEE, which can be described by
\begin{equation}\label{eq:le_alpha}
\begin{aligned}
M_{i} = \alpha^{\beta}_{ij} E_j
\end{aligned}
\end{equation}
Here $M$ and $E$ are magnetization and electric field, respectively, and $i,j = x,y,z$ are Cartesian indices. Meanwhile, $\alpha^{\beta}_{ij}$ is the LEE response tensor, and $\beta = S, L$ denotes the contributions from spin ($\beta = S$) and orbital ($\beta = L$) angular momenta. Typically, an electric (magnetic) field is used to generate an electric (magnetic) momentum. Eq.~(\ref{eq:le_alpha}), however, suggests that an electric field can induce a magnetic polarization, which is relatively less common. A critical question is: what are the necessary conditions for the emergence of such a coupling between electric and magnetic properties? We first examine the inversion symmetry $\P$. Under $\P$ operation, $E$ gets a minus sign, while $M$ remains invariant. Clearly, $\P$ must be broken for a non-zero $\alpha$. On the other hand, the requirement for time-reversal symmetry $\T$ is more subtle. Under $\T$ operation, $M$ gets a minus sign, while $E$ is invariant. Eq.~(\ref{eq:le_alpha}) thus suggest that $\T$ must also be broken for a non-zero $\alpha$. This is naturally satisfied in magnetic materials such as $\rm Cr_2 O_3$~\cite{iniguez2008first,malashevich2012full,scaramucci2012linear},  where the intrinsic magnetic moments break $\T$. In this case, the electric field generates a \emph{change} in the original magnetic moments, which is usually called the magnetoelectric effect~\cite{fiebig2005revival}.

Remarkably, $\alpha$ can be non-zero in non-magnetic metals as well, which is called the LEE~\cite{edelstein1990spin}. Since in metals, the electric field also induces a charge current, the LEE is also described as a charge-to-spin conversion process~\cite{ghiasi2019charge,el2023observation}. A non-magnetic metal, however, is time-reversal invariant, so how is $\T$ broken, which is necessary for a non-vanishing $\alpha$?  It is actually broken by dissipation and the Joule heat associated with the charge current~\cite{xu2021light}. This can be better understood by examining the charge current generation under electric fields, which can be expressed as $j_i = \sigma_{ij} E_j$ with $j$ the charge current and $\sigma$ the conductivity. Notably, $j$ gets a minus sign under $\T$ operation, so a non-zero $\sigma$ also requires the $\T$-breaking. However, the Ohm current $j_i = \sigma_{ij} E_i$ can exist in any metal. This is because the Joule heat associated with the Ohm current breaks $\T$. In contrast, the Hall current $j_i = \sigma_{ij} E_j\, (i\neq j)$  is not accompanied by heat generation and dissipation, so it can not break $\T$ by itself. Consequently, the Hall current only appears in magnetic materials~\cite{nagaosa2010anomalous}, or in the presence of external magnetic fields, which helps break $\T$.

\begin{figure}
    \centering
    \includegraphics[width=1\linewidth]{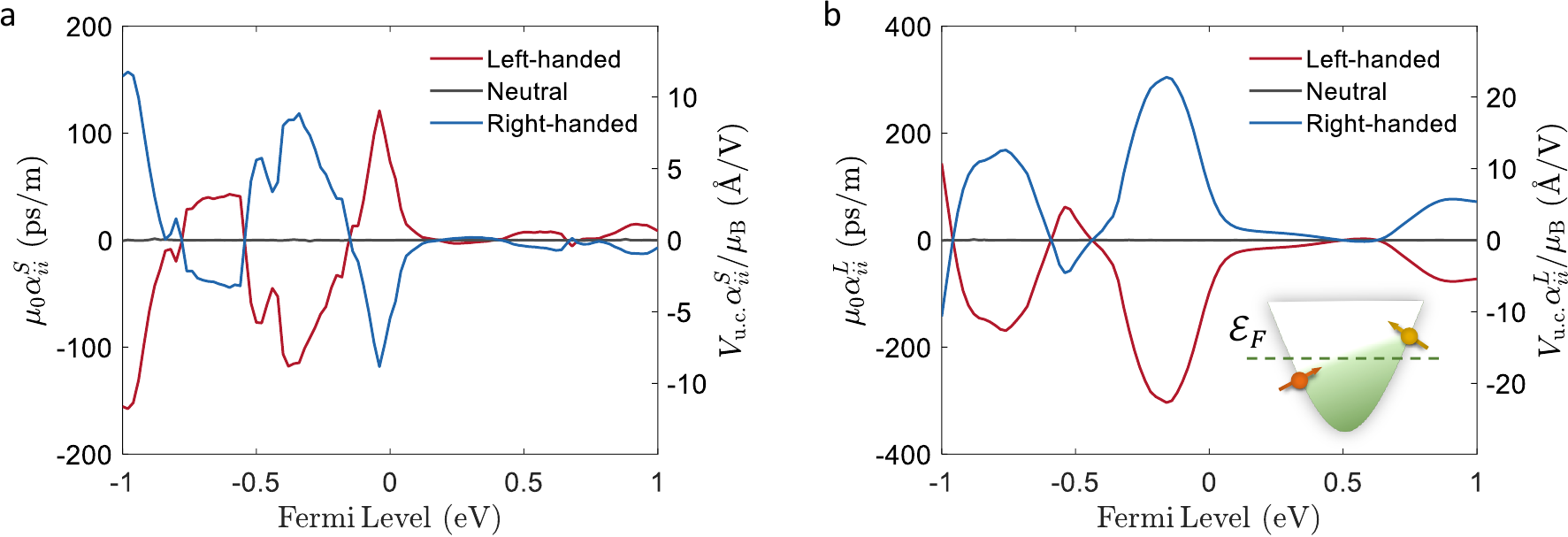}
    \caption{Linear Edelstein coefficients $\alpha^{\beta}_{ii}$ from (a) spin and (b) orbital contributions. The left $y$-axis shows $\mu_0 \alpha^{\beta}_{ii}$ in the unit of $\rm ps/m$, while the right axis shows corresponding $V_{\rm u.c.}\alpha^{\beta}_{ii}/\mu_{\rm B}$ in the unit of $\rm \AA/V$. The LEE coefficients of left-handed and right-handed CoSi are opposite to each other, while that of the neutral structure is zero due to inversion symmetry $\P$. The inset of (b) is a simplified illustration of the mechanism of LEE. Electric fields drive a change in the electron occupations around the Fermi level $\mathcal{E}_F$, and different electronic states have different spin/orbital angular momentum (arrows). Deeper (lighter) green color corresponds to larger (smaller) electron occupation numbers.}
    \label{fig:linear_edelstein}
\end{figure}

After elucidating the symmetry requirement on the LEE, we briefly discuss its microscopic mechanism.  Electron on each state $\ket{n}$ carries its magnetic momentum $\beta_n$. However, in non-magnetic materials under equilibrium, the contributions from all electrons sum up to zero. That is, one has $M \propto \sum_n f_n \beta_n=0$, where $f_n$ is the occupation number of state $\ket{n}$. When an electric field is applied, the electron occupations would be modified, and potentially one can have $M \propto \sum_{n} f^{(1)}_n \beta_n \neq 0$, where $f^{(1)}_n$ is the change in the occupation that is first order in electric field, i.e., $f^{(1)}_n \propto E$ (Inset of Figure~\ref{fig:linear_edelstein}b). The LEE response tensor, which is determined by $f^{(1)}_n$, can be obtained from the first-order perturbation theory~\cite{vzelezny2017spin,johansson2018edelstein,xu2021light}
\begin{equation}\label{eq:alpha_formula}
\begin{aligned}
\alpha^{\beta}_{ij} = \frac{i e \mu_{\rm B} }{\hbar} \int \frac{\d k}{(2\pi)^3}
&  \left[ \sum_{m\neq n} \frac{f_{nk}-f_{mk}}{\omega_{mk} - \omega_{nk}} \frac{\bra{nk} \beta_i \ket{mk} \bra{mk}  v_j \ket{nk} }{\omega_{mk} - \omega_{nk} + i\delta} \right. \\
& - \left. \sum_n \frac{\partial f_{nk}}{\partial \omega_{nk}} \frac{\bra{nk} \beta_i \ket{nk} \bra{nk}  v_j \ket{nk}}{i\delta}
\right]
\end{aligned}
\end{equation}
Here, the first (second) term in the bracket is the interband (intraband) contribution to the total LEE response. $\mu_0$, $e$, and $\hbar$ are the vacuum permeability, electron charge, and the reduced Planck constant, respectively. $\ket{nk}$ is the Bloch state at band $n$ and wavevector $k$, whose occupation number and band energy are $f_{nk}$ and $\hbar \omega_{nk}$, respectively. The derivative $\frac{\partial f_{nk}}{\partial \omega_{nk}}$ is non-zero near the Fermi surface in metals. In semiconductors or insulators at zero temperature, one has $\frac{\partial f_{nk}}{\partial \omega_{nk}}\equiv 0 $, and the intraband contribution is thus absent. Meanwhile, $v$ is the velocity operator, and $\beta=2S$ ($\beta = L$) is the spin (orbital) angular momentum operator. The additional factor of 2 for spin accounts for the electron $g$-factor. $\delta$ is a smearing factor that corresponds to the finite linewidth of the electronic states, and we take a uniform value of $\hbar \delta = 20~\rm meV$, corresponding to an electron lifetime of around $0.2~\rm ps$.

Next, we demonstrate the LEE in CoSi, a chiral topological semimetal. The atomic structure of CoSi (Figure~\ref{fig:atomic_band_structure}a,~b) has a chiral space group of $\rm P2_1 3$ (no.~198). Along the $(111)$ direction, the Co (Si) atoms form helical structures that spiral clock-wisely (anti-clock-wisely), which will be denoted as the left-handed structure. The mirror image, whereby the atoms spiral in opposite directions, corresponds to the right-handed structure. The left- and right-handed structures are non-superimposable mirror (inversion) images of each other, a characteristic known as chirality~\cite{yan2024structural}.  Due to the cubic symmetry of CoSi, the only non-zero component~\cite{gallego2019automatic} of $\alpha$ is  $\alpha_{ii} \equiv \alpha_{xx} = \alpha_{yy} = \alpha_{zz}$, which is plotted against the Fermi level in Figure~\ref{fig:linear_edelstein}. The Fermi level is offset to its charge neutral value, i.e., the zero point in Figure~\ref{fig:atomic_band_structure}b. To make a more convenient comparison with standard experimental definitions of magnetoelectric responses~\cite{malashevich2012full}, we show $\mu_0 \alpha_{ii}$ in the unit of $\rm ps/m$ (left $y$-axis in Figure~\ref{fig:linear_edelstein}). Note that here we assume the permeability of CoSi is $\mu = \mu_0$, which is a good approximation since CoSi is non-magnetic. One can see that $\mu_0 \alpha_{ii}$ can be as large as $10^2~\rm ps/m$, which is about two-orders of magnitude than the magnetoelectric tensor of $~\rm Cr_2O_3$~\cite{iniguez2008first,malashevich2012full,kita1979experimental,wiegelmann1994magnetoelectric}, though still smaller than that of certain two-phase systems, which can reach $10^3~\rm ps/m$~\cite{eerenstein2006multiferroic,ryu2001magnetoelectric,cai2004large}. It is also useful to directly calculate how much magnetic moment per unit cell can be generated by the electric field. Hence, we also plotted $V_{\rm u.c.} \alpha_{ii}/\mu_{\rm B}$ in the unit of $~\rm \AA/V$  (right $y$-axis in Figure~\ref{fig:linear_edelstein}), where $V_{\rm u.c.}$ is the volume of the CoSi unit cell.  The magnetic momentum induced by the electric field is on the order of $10~\rm \mu_{\rm B}$ per unit cell with $E = 1~\rm V/\AA$. This is reasonable - the intrinsic atomic electric fields are typically on the order of $E_{\rm in} \sim 1~\rm V/\AA$.  ($E_{\rm in}$ is the electric field strength generated by electrons and nuclear charge  near the nucleus). When $E$ is comparable to $E_{\rm in}$, the electronic states would be significantly modified, which can result in a magnetic moment comparable to that in intrinsic magnetic materials. Nevertheless, it should be noted that Eq.~(\ref{eq:alpha_formula}) comes from first-order perturbation theory, and it applies to any electric field strength that satisfies $E\ll E_{\rm in}$.  When $E\sim E_{\rm in}$, non-perturbative approaches should be used.

Interestingly, the LEE of left- and right-handed CoSi (two enantiomers) are opposite to each other because their structures are images related by inversion symmetry. From Eq.~(\ref{eq:le_alpha}), one can see that $\alpha$ acquires a minus sign after a $\P$ operation is applied. Furthermore, for the ``neutral" structure (space group $\rm Fm\bar{3}m$, no.~225), which is positioned symmetrically between the left- and right-handed structures, the LEE is forbidden due to its inversion symmetry. This highlights the importance of inversion symmetry in the LEE.

\section{Nonlinear Edelstein effect}

Next, we turn to NLEE, by which magnetization can be generated with light illumination~\cite{xu2021light}. Basically, the NLEE can be described by
\begin{equation}\label{eq:nlee_response}
\begin{aligned}
M_{i} = \sum_{\Omega = \pm \omega} \chi^{\beta}_{ijk} E_j(\Omega) E_k(\Omega)
\end{aligned}
\end{equation}
 where $\chi^{\beta}_{ijk}$ is the NLEE response tensor. Eq.~(\ref{eq:nlee_response}) indicates that the $\omega$ and $-\omega$ components of the light field are combined, and a static magnetization is generated. This is similar in essence to other second-order nonlinear optical effects, such as the bulk (spin) photovoltaic, where the light illumination generates charge (spin) current~\cite{xu2021pure,dai2023recent}. We also need to analyze the symmetry requirements on the NLEE. Notably, $\chi$ is invariant under inversion symmetry operation as $E$ ($M$) gets a $-1$ $(+1)$ sign under $\P$ operation. Hence, the NLEE does \emph{not} need inversion symmetry breaking. This should be distinguished from many other well-known second-order nonlinear optical effects, including bulk (spin) photovoltaic effect and second harmonics generation, which require inversion symmetry breaking. Regarding time-reversal symmetry, it is straightforward to show that NLEE requires $\T$ breaking, as in the case of LEE. However, light illumination naturally generates heat and dissipation, which spontaneously breaks $\T$ symmetry, according to the second law of thermodynamics. As a result, NLEE can exist in non-magnetic materials under linearly polarized light, as shown below.

\begin{figure}
    \centering
    \includegraphics[width=1\linewidth]{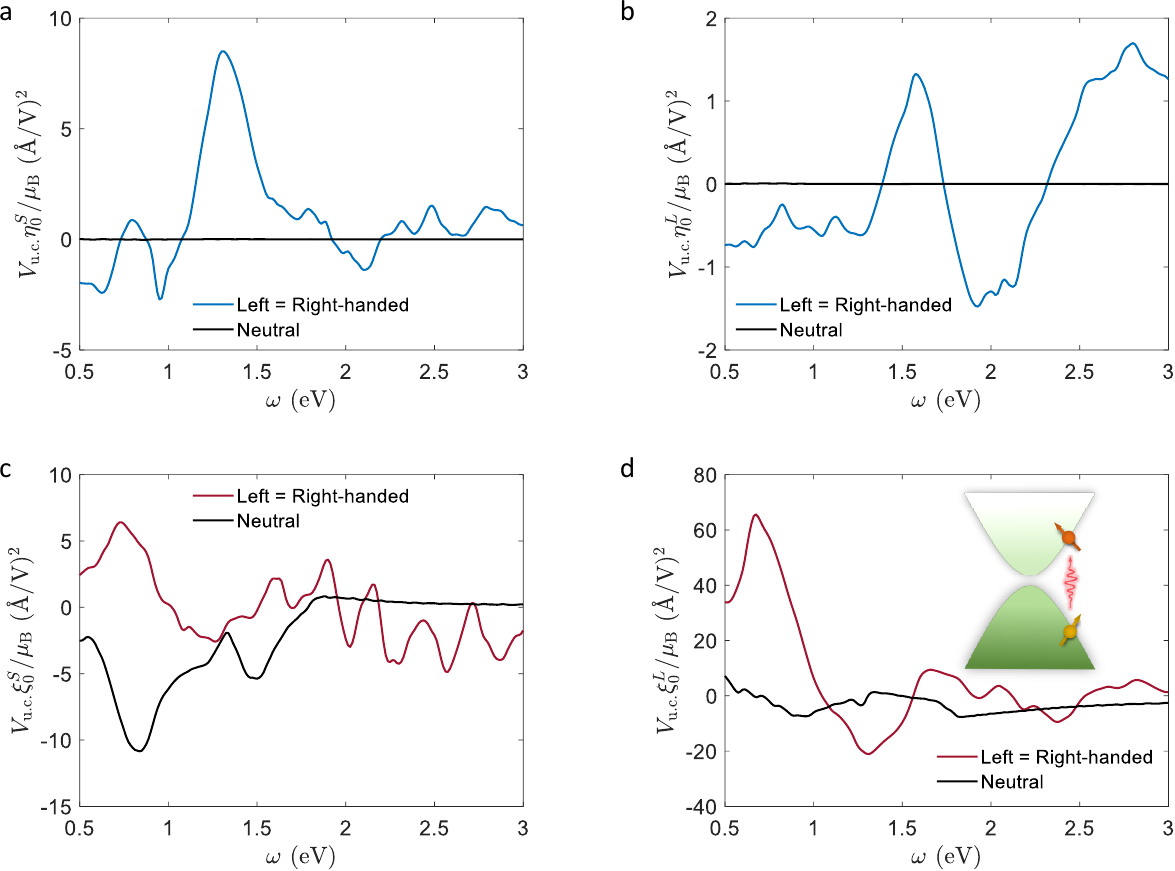}
    \caption{Nonlinear Edelstein coefficients under (a, b) linearly polarized light and (c, d) circularly polarized light. (a, c) and (b, d) are contributions from spin and orbital angular momenta, respectively. The NLEE coefficients of the left- and right-handed CoSi are the same, while $\eta^{\beta}$ of the neutral structure is zero due to its higher symmetry. Inset of (d) shows a simplified illustration of the mechanism of the NLEE. Light (vertical wavy arrow) drives interband transitions, resulting in a change in the electron occupations. The deeper (lighter) green color corresponds to larger (smaller) electron occupation number. The spin/orbital angular momentum of the electronic states are denoted by the arrows. }
    \label{fig:nonlinear_edelstein}
\end{figure}

Microscopically, light illumination leads to a static change in the electron occupation $f^{(2)}_n$ that is second order in the electric field, that is, $f^{(2)}_n \propto E(\omega) E(-\omega)$. Then, the NLEE magnetization is $M \propto \sum_n f^{(2)}_n \beta_n$, where $\beta_n$, as we introduced before, is the magnetic moment carried by electronic state $\ket{n}$. Using the second-order perturbation theory, one has~\cite{xu2021light,xu2021pure}
\begin{equation}\label{eq:chi}
\begin{aligned}
\chi^{\beta}_{ijk} (\omega) = -\frac{\mu_{\rm B} e^2}{\hbar^2 \omega^2} \int \frac{\d k}{(2\pi)^3} & \sum_{mnl} \frac{(f_{lk}-f_{mk})\bra{lk}v_j\ket{mk}}{\omega_{mk} - \omega_{lk} -\omega + i\delta} \\
& \times \left[
\frac{\bra{mk}\beta_i\ket{nk} \bra{nk}v_k \ket{lk}}{\omega_{mk}-\omega_{nk} + i\delta}
-\frac{\bra{mk} v_k \ket{nk}  \bra{nk}\beta_i\ket{lk}}{\omega_{nk}-\omega_{lk} + i\delta}
\right]
\end{aligned}
\end{equation}
Note that the $k$ subscripts in $\chi^{\beta}_{ijk}$ and $v_k$ are Cartesian indices, while others denote the electron wavevector. In order to examine the response under linearly and circularly polarized light, we define~\cite{xu2021light}
\begin{equation}\label{eq:eta_xi}
\begin{aligned}
\eta^{\beta}_{ijk} & = \frac{1}{2}\Re \left\{ \chi^{\beta}_{ijk} +  \chi^{\beta}_{ikj} \right\} \\
\xi^{\beta}_{ijk} & = \frac{1}{2}\Im \left\{ \chi^{\beta}_{ijk} -  \chi^{\beta}_{ikj} \right\}
\end{aligned}
\end{equation}
Here $\eta$ is the symmetric real part of $\chi$, which is the response under linearly polarized light. Meanwhile, $\xi$ is the anti-symmetric imaginary part of $\chi$, corresponding to circularly polarized light. With a space of $\rm P2_13$, it can be show that both $\eta^{\beta}_{ijk}$ and $\xi^{\beta}_{ijk}$ have only one independent element~\cite{gallego2019automatic}. Specifically, one has $\eta^{\beta}_{ijk} = \eta^{\beta}_0 \vert \varepsilon_{ijk} \vert$ and $\xi^{\beta}_{ijk} = \xi^{\beta}_0 \varepsilon_{ijk}$, where $\varepsilon_{ijk}$ is the Levi-Civita symbol. The calculated results of $\eta^{\beta}_0$ and $\xi^{\beta}_0$ are plotted in Figure~\ref{fig:nonlinear_edelstein} for both spin and orbital contributions. Several peaks can be observed, and they arise (1) when there is a peak in the joint density of states between the valence and conduction bands (somewhat similar to the van Hove singularity in electron density of states), and/or (2) when the interband transition dipoles reach maximum at certain light frequencies.  Multiply by $V_{\rm u.c.}/\mu_{\rm B}$, the tensors have a unit of $\rm (\AA/V)^2$. One can see that the light-induced magnetic moments per unit cell can be on the order of  $1\sim 10~\mu_{\rm B}$ when the light field is $E=1~\rm V/\AA$. This phenomenon can be understood in a manner similar to LEE: the light-induced magnetic moment becomes comparable to that of magnetic materials when the electric field of the light approaches the strength of the intrinsic atomic electric field.

Different from the LEE, the left- and right-handed CoSi exhibit identical NLEE responses. Indeed, applying a $\P$ operation on Eq.~(\ref{eq:nlee_response}), one can see that $\chi$ remains the same. Hence, the NLEE  of a material would be identical to that of its inversion image. In some literature, this phenomenon is called non-reciprocal~\cite{tokura2018nonreciprocal}. Interestingly, the NLEE can exist in the achiral neutral structure, although it is inversion symmetric. The symmetric real part $\eta^{\beta}$, however, must be zero due to the high symmetry of the neutral structure (space group $\rm Fm\bar{3}m$, no.~225)~\cite{gallego2019automatic}. Note that generally, both $\eta^{\beta}$ and $\xi^{\beta}$ can be non-zero in inversion symmetric materials with lower symmetries.

\section{Conclusions}

In conclusion, we present a quantum theory for the linear and nonlinear Edelstein effects (LEE and NLEE), demonstrating how static magnetization can be generated through electric and optical fields, respectively. Using first-principles calculations, we illustrate the presence of LEE and NLEE in the chiral topological semimetal CoSi. Both effects are applicable across a wide range of materials and hold promise for various applications, including magnetic memory and spintronics.

\section{Acknowledgments}
This work was supported by the Office of Naval Research Multidisciplinary University Research Initiative Award No. ONR N00014-18-1-2497.

\appendix

\section{Methods}

 The first-principles calculations in this work are based on the density functional theory (DFT)~\cite{hohenberg1964inhomogeneous,kohn1965self} as implemented in the Vienna \textit{ab initio }simulation package (VASP)~\cite{kresse1996efficiency,kresse1996efficient}. Generalized gradient approximation (GGA) in the form of Perdew-Burke-Ernzerhof (PBE)~\cite{perdew1996generalized} is used to approximate the exchange-correlation interactions among electrons. Core electrons are treated by the projector augmented wave (PAW) method~\cite{blochl1994projector}, while valence electrons wavefunctions are expanded by plane-waves. For \emph{ab initio} DFT calculations, the first Brillouin zone is sampled by a $\Gamma$-centered $9\times9\times9$ $k$-mesh for CoSi. Tight-binding Hamiltonians in real space atomic basis are subsequently generated from the Bloch Hamiltonian in DFT calculations, using the Wannier90 package~\cite{mostofi2014updated}. The tight-binding Hamiltonian is then used to interpolate the band structure, the $v$, $S$, and $L$ matrices on a much denser $k$-mesh ($128\times 128\times 128$ for CoSi) to calculate the LEE and NLEE response functions. The orbital angular momentum $L$ is calculated using the angular momentum matrices of localized atomic orbitals (s, p, d, etc.), which only contain the ``intra-atom'' contribution $\bra{nR} r\times p \ket{n'R}$, where $\ket{nR}$ is the $n$-th orbital of the atom at location $R$,  while $r$ and $p$ are position and momentum operators, respectively.  That is, if the orbital baisis set is $\{s, p, d\}$, then the angular momentum operator is defined as $L=L_s \bigoplus L_p \bigoplus L_d$, where, for example, $L_d$ is the angular momentum operator in the subspace of $d$-orbitals. Meanwhile, the ``inter-atom" contribution $\bra{nR} r\times p \ket{n'R'} \, (R'\neq R)$ is neglected. A more rigorous treatment of the orbital angular momentum may be realized by using e.g., the modern theory of orbital magnetization~\cite{malashevich2010theory,thonhauser2005orbital,shi2007quantum,essin2009magnetoelectric}.

\bibliography{bibliography}

\end{document}